\documentstyle[prb,aps,multicol,epsf]{revtex}
\newcommand{\startlongequation}{
\end{multicols}\vspace*{-3.5ex}{\tiny\noindent
\begin{tabular}[t]{c|} \parbox{0.493\hsize}{~} \\ \hline \end{tabular}} }
\newcommand{\stoplongequation}{
{\tiny\hspace*{\fill}
\begin{tabular}[t]{|c}\hline\parbox{0.49\hsize}{~} \\ \end{tabular}}
\vspace*{-2.5ex}\begin{multicols}{2}}
\begin{document}
\draft
\title{A self-interaction corrected pseudopotential scheme for magnetic
and strongly-correlated systems.}
\author{Alessio Filippetti and Nicola A. Hill}
\address{Materials Department, University of California,
Santa Barbara, CA 93106-5050}
\maketitle

\begin{abstract}
Local-spin-density functional calculations may be affected by severe errors
when applied to the study of magnetic and strongly-correlated materials.
Some of these faults can be traced back to the presence of the spurious
self-interaction in the density functional. Since the application 
of a fully self-consistent self-interaction correction is highly demanding 
even for moderately large systems, we pursue a strategy of approximating 
the self-interaction corrected potential with a non-local, pseudopotential-like 
projector, first generated within the isolated atom and then updated during the 
self-consistent cycle in the crystal. This scheme, whose implementation is totally
uncomplicated and particularly suited for the pseudopotental formalism, 
dramatically improves the LSDA results for a variety of compounds with a minimal 
increase of computing cost.
\end{abstract}

\begin{multicols}{2}
\section{Introduction}
\label{introd}
Among the innumerable successes of the local-density (LDA) and local-spin-density (LSDA) 
approximations to density functional theory (DFT) there are also well known, systematic 
failures\cite{jones,jones1} that compromise the accuracy of predictions for a range of 
properties, especially in magnetic and strongly-correlated\cite{note_cor} compounds. 

Typical examples of LSDA failures are the series of transition-metal monoxides,\cite{fulde} 
improperly described in LSDA as either small-gap Mott-Hubbard antiferromagnetic insulators 
(MnO, NiO)\cite{towk,sa} 
or even ferromagnetic and non-magnetic metals (FeO, CoO, CuO),\cite{towk,zscrhbd} whereas according
to experiments these materials are charge-transfer antiferromagnetic wide-gap insulators. 
A similar situation occurs for the high-T$_c$ compounds La$_2$CuO$_4$ and YBa$_2$Cu$_3$O$_6$ 
which are described in LSDA as nonmagnetic metals instead of as antiferromagnetic insulators,\cite{pickett}
and for the perovskite manganites (e.g. La$_x$Ca$_{1-x}$MnO$_3$),\cite{ps,smth} for which the 
LSDA fails to predict the correct magnetic and orbital orderings. In general, LSDA favors metallic 
and ferromagnetic ground states over the observed antiferromagnetic insulating ground states.
This is particularly harmful in the case of hexagonal YMnO$_3$, which is antiferromagnetic and 
ferroelectric, but is described as a metal within LSDA,\cite{fh} 
thus preventing the possibility of calculating any ferroelectric properties at all.

These failures can be, at least in part, attributed to the presence of the self-interaction 
(SI) in the LSDA energy functional, i.e. the interaction of an electron charge with the 
Coulomb and exchange-correlation potential generated by the same electron.
The SI vanishes in the thermodynamic limit for delocalized states, but is present in systems 
characterized by spatially localized electron charges like 2p, 3d and 4f 
electrons. As a consequence of the SI, the binding energies, the on-site Coulomb energies 
(i.e. the Hubbard U parameter) and the exchange-splitting of the d and f states are underestimated, 
whereas the hybridizations of cation d and anion p states and the corresponding band widths are 
overemphasized.

Since, for materials with partially filled d states, the on-site Coulomb energy should be 
much larger than the charge-transfer energy between p and d electrons, it is clear that 
the suppression of U and the overestimation of the p-d hybridization may change dramatically  
the character of the band structure, with a tendency to favor metallic and 
ferromagnetic ground states over insulating antiferromagnetic states.
Furthermore, within the LSDA Kohn-Sham (KS) description based on a local, orbital-independent 
potential, orbital and charge orderings cannot be properly accounted for.

The presence of SI and the possible strategies for eliminating the SI in density functional 
theories are long-standing issues which go back to the Thomas-Fermi and the Slater (X$\alpha$) 
approaches.\cite{history} A large amount of literature has been produced\cite{zung,perd,zpo,gj,pz} 
within the context of the LSDA. Particularly fundamental are the works by 
Perdew and Zunger\cite{zpo,pz} which proposed a form of self-interaction correction (SIC) within 
LSDA (SIC-LSDA) and successfully applied it to the calculation of atomic properties. 

More recently, Svane and co-workers presented a long series of 
works\cite{sg,svane,aka,stw,svane1,btsg,ls,ls1,stspsw,twss,psts} which attempted
the highly challenging application of the fully self-consistent SIC-LSDA to extended 
systems with encouraging results.
However the full SIC-LSDA requires a large computing effort when applied to extended 
systems even for materials with small unit cells, and becomes prohibitive for larger systems.

A useful alternative to this approach has been suggested by Vogel and 
co-workers.\cite{vkp,vkp1,vkp2} They approximated the SI part of the KS potential 
with a non-local, atomic-like contribution included within the pseudopotential construction.
This scheme, applied to non-magnetic II-VI semiconductors and III-V nitrides, is capable of 
remarkable improvements upon the LSDA results while keeping the computational cost comparable to 
that of an ordinary pseudopotential calculation.

Inspired by these results, in this paper we develop an approach which is in the same 
spirit as that of Ref.\onlinecite{vkp}, but can be applied to more general cases, and in particular 
to magnetic and highly-correlated systems where there is a coexistence of strongly localized and 
hybridized electron charges. The main innovation of our formalism 
is the introduction of orbital occupation numbers in the electron KS Hamiltonian: The pseudopotential-like 
SIC (pseudo-SIC) is rescaled by the occupation numbers calculated self-consistently within 
the crystal environment, so that the SI coming from localized, hybridized, or completely itinerant 
states may be discriminated accordingly. Also, these occupation numbers discriminate between valence and 
conduction bands so that only the former are corrected, in accord with the idea that conduction states 
are itinerant, thus SI-free. Furthermore, our scheme is generally applicable to insulators as well
as metals, without the necessity of knowing the character of the material a priori.
This is essential for compounds which are erroneously described as metals within LSDA, such as the 
hexagonal YMnO$_3$, treated in section \ref{res3}.

A second major difference between our approach and Ref.\onlinecite{vkp} lies in the way in which relaxation 
corrections to the SIC potential are accounted for. In Ref.\onlinecite{vkp} they are incorporated as an 
additional, orbital-dependent number which is imported from atomic energy difference calculations 
(i.e. the so-called $\Delta$SCF method). In contrast, in our case the relaxation contribution is included 
directly as an analytical correction in the SIC potential projector, and no further atomic calculations 
are needed. 

In this paper we test the pseudo-SIC on three classes of materials, including the wide-band gap 
insulators (e.g. ZnO and GaN), the transition-metal oxides (MnO and NiO) and the manganese 
oxides (YMnO$_3$). Our main focus is on the electronic properties (i.e. the band energy structure) 
of these materials, since we want to verify the capability of our single-electron Hamiltonian to 
reproduce the main features of photoemission spectra. For some of these compounds we also calculated
the theoretical structure by total energy minimization. In general we obtain very encouraging results 
and clear systematic improvements over the LSDA description, without considerable increase of 
computing effort.

Note that we have implemented the pseudo-SIC within the ultrasoft pseudopotential method\cite{van}
(USPP). This is instrumental in keeping a moderate computational cost even for large unit cells
containing atoms with 2p and 3d electrons.

As is customary when a new scheme is introduced, we compare our results with those of 
other common beyond-LDA approaches. According to the set of results presented here, the pseudo-SIC 
ranks among the most accurate. In particular, it seems to perform equally well (or even better) 
than the very popular LDA+U.\cite{aza} Although a detailed comparison between pseudo-SIC and LDA+U 
is not the aim of this paper, we can point out some potential advantages for the pseudo-SIC.
First, it does not require parameters imported from an external theory. Second, it can be applied 
to both magnetic and non-magnetic compounds, whereas the LDA+U is constructed to correct 
spin-polarized and/or orbital-ordered band structures. 

Finally, within LDA+U or even within the exact SIC-LSDA a choice of which orbitals are 
localized in space and thus, which orbitals are to be corrected, has to be made before the 
calculation, and the final result depends on this assumption. 
Instead, in pseudo-SIC the correction is applied to all the orbitals indiscriminately, 
and no choice of orbital localization is required. This is an advantage in case where a discrimination 
of the electron charge localization cannot be established a priori. This happens in non-bulk 
systems such as surfaces and interfaces, when localized surface states or resonance states are present, 
or even in bulk materials whenever the charge localization is not, or not only, due to the 
chemical nature of the involved electrons. Examples of this kind are the superconductor cuprates, 
in which the d orbital charges have different localizations corresponding to different 
physical solutions.

The remainder of this paper is organized as follows: in Section \ref{overv} we review the main 
features of earlier work on SIC implementations in LSDA.
In Section \ref{method} we describe the pseudo-SIC formulation applied in combination 
with norm-conserving pseudopotentials (NCPP). The extension of the pseudo-SIC to USPP is 
given in the Appendix. In Section \ref{results} we report our results obtained within the 
pseudo-SIC approach, and finally in Section \ref{conc} we present the conclusions.

\section{Overview of previous work}
\label{overv}

The SI effects in {\it atomic} calculations are extensively explained in the seminal paper by
Perdew and Zunger.\cite{pz} In the following we briefly summarize the most fundamental features.
Due to the SI, the tail of the KS potential does not recover its physical long-range limit, -1/$r$. 
Thus negative ions which are stable experimentally cannot be described at all in LSDA
since the outermost eigenstates are unbound. The atomic total energies are severely 
overestimated with respect to the experimental values, as a consequence of the reduction in 
binding energy caused by the spurious self-screening in the KS potential.

Consistently, the LSDA KS eigenvalues strongly overestimate the experimental electron
removal energies. Although the KS eigenvalues represent in general Lagrange multipliers
and the comparison with the removal energies may be questioned, calculations\cite{pz}
clearly show that in atoms the SI represents, by far, the largest source of mismatch.
Furthermore, it can be proved that the highest occupied KS eigenvalue of the exact density
functional theory equals the atomic ionization potential.\cite{pplb} Thus, at least for this 
quantity, the discrepancy must be attributed to the LSDA inaccuracy.

The SIC-LSDA proposed by Perdew and Zunger is based on the straightforward subtraction 
of the SI contribution from the LSDA KS potential:

\begin{eqnarray}
V_{HXC}^{\sigma}[n,m] \rightarrow V_{HXC}^{\sigma}[n,m] - V_{HXC}^{\sigma}[n_i^{\sigma}],
\label{eq1}
\end{eqnarray}

where $n$ and $m$ are the total charge and magnetization densities, and $n_i^{\sigma}$ is the 
spin-charge density of the $i^{th}$ orbital. It is understood that in $V_{HXC}^{\sigma}[n_i^{\sigma}]$
the magnetization density is set equal to 1, i.e. the single electron charge is fully
spin-polarized. Despite its conceptual simplicity, this modification is able to systematically 
improve total energies, ionization potentials and electron affinities, giving a much better agreement 
between KS eigenvalues and removal or addition energies.

Notwithstanding this initial success, the SIC-LSDA has not enjoyed a large popularity in the
first-principles community, mainly because the correction severely compromises
the feasibility of the LSDA, introducing an orbital-dependent, spatially localized
(i.e. non-periodic) contribution in the KS potential. Thus different wavefunctions experience 
different Hamiltonians, and are no longer orthogonal.
This is not a serious drawback in atomic calculations, since the non-orthogonality is a small 
effect and the atomic orbitals can be easily forced to be orthogonal during the 
self-consistent cycle.\cite{pz} 

In contrast, the direct application of Eq.\ref{eq1} to extended systems (where the solutions
of the KS equations are Bloch states) is very challenging. First of all the SI of a Bloch 
state is an ill-defined quantity that depends on the normalization of the wavefunction, 
and vanishes as $\Omega^{-1/3}$ in the thermodynamic limit, where $\Omega$ is the system volume. 
Thus, if we assume that the electron charges are properly described as Bloch states 
the SIC becomes immaterial and discardable.

Clearly this is not necessarily true in general. In cases like simple metals or semiconductors 
with mostly covalent interactions (e.g. bulk Si) we can safely assume that the SI is discardable. 
However, in many systems the electron charges retain atomic-like features such as small band 
dispersion and a spatial distribution strongly localized around their ion-cores. 
A clear example of the variable influence of the 
SI is given by the energy gap of semiconductors: for bulk Si the discrepancy between the theoretical 
and experimental energy gaps (0.7 eV and 1.17 eV, respectively)  must be attributed to non-locality 
and many-body effects. Instead, in LiF the SIC accounts for almost 95\% of the LDA gap error.\cite{zf}

Thus, in order to have a non trivial SIC, a description in terms of localized orbitals must be 
adopted, which causes violation of the translational invariance and consequently the inapplicability 
of the Bloch theorem. 
Furthermore, the SIC-LSDA eigenstates are not invariant under unitary rotations within the subspace 
of the occupied orbitals, and the SIC-LSDA solutions strictly depend on the assumption of choosing 
each orbital as extended (thus self-interaction free) or localized (therefore subject to SIC). 
We will show an example of this in Section \ref{res2}.

To our knowledge the implementation of a fully self-consistent SIC-LSDA approach for extended 
systems was pioneered by Svane and co-workers.\cite{sg} They carried out a series of applications 
to a remarkable range of materials including the family of transition-metal monoxides,\cite{sg,ls} 
the high-T$_c$ superconductor parent compounds La$_2$CuO$_4$\cite{svane,aka} and 
YBa$_2$Cu$_3$O$_6$,\cite{twss} the rare-earth materials $\gamma$-Ce and 
$\alpha$-Ce\cite{svane1,btsg,ls1}, and Yb\cite{stspsw} and Pu\cite{psts} 
monopnictides and monochalcogenides, obtaining systematic improvements over the LSDA results.
Other SIC-LSDA applications to transition-metal oxides,\cite{stw,af} implemented within 
different computational methodologies, are also present in the literature.

The major drawbacks of these SIC-LSDA implementations are the rather complicated formulation 
with respect to the LSDA and the increase of computing cost which makes the SIC-LSDA 
almost impractical for large systems. One reason for this increase is the use of big supercells 
needed to describe the localized orbitals (e.g. $\sim$ 500 atoms for bulk MnO\cite{svane,af}). 
Indeed, in the works previously cited, the SIC-LSDA is implemented using a basis of linear muffin 
tin orbitals within the atomic sphere approximation (LMTO-ASA), whereas, to our knowledge, there 
are no examples of implementations using the more expensive plane-wave basis and pseudopotentials.

An important step towards a practical (albeit approximate) expression for the SIC-LDA was 
accomplished by Vogel and co-workers\cite{vkp} by incorporating the atomic SIC within the 
non-local pseudopotential projectors (SIC-PP) generated from the free atom. The idea underlying 
this approach is that the SI potential of a localized electron in the crystal can be well 
approximated by the SI which the same electron experiences in the free atom. The SIC-PP 
turned out to be quite efficient for describing the properties of some highly ionic 
compounds with atomic-like, poorly hybridized bands, such as II-VI semiconductors,\cite{vkp} 
III-V nitrides,\cite{vkp1} and silver halides.\cite{vkp2} For these materials, the energy 
band structures calculated with the SIC-PP show a much better agreement with photoemission 
spectroscopy measurements than the LDA band structure calculations. 

Our pseudo-SIC can be considered a generalization of the SIC-PP approach. It is still 
based on the idea of replacing the SIC potential with a non-local projector, but now the projector 
depends on the orbital occupation numbers calculated self-consistently within the crystal.
The details are described in the next section. 

\section{Formulation of the Pseudo-SIC}
\label{method}

\subsection{Kohn-Sham equations within pseudo-SIC}
\label{method1}

Although the calculations reported in this paper have been performed within the USPP method,\cite{van}
here we describe the pseudo-SIC formalism adapted to NCPP. This allows a simpler formulation and an 
easier understanding of the logic behind this approach. 
(The generalization to USPP is given in the Appendix.)

Within pseudo-SIC, the SIC potential is cast in terms of a non-local projector, which resembles 
the non-local part of the pseudopotential:

\begin{eqnarray}
V_{HXC}^{\sigma}[n,m] \rightarrow V_{HXC}^{\sigma}[n,m] -\sum_i\>
|{\cal Y}_i\rangle \> V_{HXC}^{\sigma}[n_i^{\sigma}]\>\langle {\cal Y}_i|.
\label{sic}
\end{eqnarray}

Here $n$ and $m$ are the periodic charge and magnetization densities of the crystal and 
the SI is written in terms of atomic quantities only: $i=[(l_i,m_i),{\bf R}_i]$ is a cumulative 
index for angular momentum quantum numbers and atomic coordinates, ${\cal Y}_i$ are projector functions 
(e.g. spherical harmonics), and $n_i^{\sigma}$ the charge densities of the (pseudo) 
atomic orbitals $\phi_i$:

\begin{eqnarray}
n_i^{\sigma}({\bf r})\,=
p_i^{\sigma}\>|\phi_i({\bf r})|^2,
\label{n_i}
\end{eqnarray}

where $p_i^{\sigma}$ are orbital occupation numbers.
Through Eq.\ref{sic}, the Bloch wavefunctions are projected onto the basis of the atomic
orbital charges $n_i^{\sigma}$. For each projection, the Bloch state is corrected by an amount 
corresponding to the atomic SIC potential $V_{HXC}^{\sigma}[n_i^{\sigma}]$. Thus the
SIC is applied without really introducing a dependence of the KS Hamiltonian on the 
individual Bloch wavefunctions, and the difficulties of the SIC-LSDA approach are overcome.

The presence of $p_i^{\sigma}$ in Eq. \ref{n_i} is a major novelty of our approach. In 
Ref.\onlinecite{vkp} the occupation numbers are implicitly set to 1, i.e. the SIC potentials 
are generated from fully occupied atomic states. Clearly, this cannot be a correct choice in 
general, since the occupation numbers may become fractions whenever hybridization, degeneracy, 
or spin-polarization effects arise. Furthermore, moving from the free atom to the crystal, 
the $p_i^{\sigma}$ can change a great deal. For instance atomic orbitals which are fully occupied 
in the atomic ground state may become conduction states in the crystal (e.g. Zn 4s in ZnO).

Thus the $p_i^{\sigma}$ must be allowed to be fractional and must be recalculated self-consistently 
within the crystal. Within a plane-wave basis set it is straightforward to calculate $p_i^{\sigma}$ 
as atomic orbital projections onto the manifold of the occupied Bloch states:

\begin{eqnarray}
p_i^{\sigma}=\,\sum_{n{\bf k}}\, f_{n{\bf k}}^{\sigma}\, \langle\psi_{n{\bf k}}^{\sigma}|
\phi_i\rangle \> \langle\phi_i|\psi_{n{\bf k}}^{\sigma}\rangle.
\label{occ}
\end{eqnarray}

These quantities are analogous to the local orbital occupations calculated within the FLAPW implementation 
of the LDA+U.\cite{slp,babh} In the limit of an isolated atom the $p_i^{\sigma}$ recover the atomic values, 
whereas in the case of hybridized bonds, they rescale the atomic SIC by the amount of charge that actually 
occupies the atomic orbital. This fact has a fundamental consequence: since in most cases empty bands have 
dominant characters from orbitals whose occupation numbers are close to zero, the SIC potential will not 
affect these bands. This is consistent with the general assumption that conduction states are itinerant, 
thus SI-free. (In fact, even if the conduction states are not itinerant, it is theoretically justified to 
apply the SIC only to the occupied states, since these are the states which actually see their own 
charge.\cite{comm1}) Notice that partially occupied bands can be treated on the same footing as filled bands, 
thus the scheme can be applied to both insulators and metals. 

Following the suggestion of Ref.\onlinecite{vkp}, for the purpose of numerical efficiency 
we recast the pseudo-SIC as a fully non-local, Kleinman-Bylander projector: 

\begin{eqnarray}
\hat{V}_{SIC}^{\sigma}\,=\,\sum_i\> 
{|\gamma_i^{\sigma}\,\rangle\,\langle\,\gamma_i^{\sigma}|\over C_i^{\sigma}},
\label{sic1}
\end{eqnarray}

where
\begin{eqnarray}
\gamma_i^{\sigma}({\bf r})\, =\,V_{HXC}^{\sigma}[n_i^{\sigma}({\bf r})]
\,\phi_i({\bf r})
\label{gamma}
\end{eqnarray}

and
\begin{eqnarray}
C_i^{\sigma}=\langle\phi_i|V_{HXC}^{\sigma}[n_i^{\sigma}]|\phi_i\rangle.
\label{c_i}
\end{eqnarray}

The pseudo-SIC KS equations are:

\begin{eqnarray}
\left[-\nabla^2 + \hat{V}_{PP} + \hat{V}_{HXC}^{\sigma} 
- \hat{V}_{SIC}^{\sigma}\right]|\psi^{\sigma}_{n{\bf k}}\rangle=
\epsilon^{\sigma}_{n{\bf k}}\,|\psi^{\sigma}_{n{\bf k}}\rangle,
\label{ks-sic}
\end{eqnarray}

where $\hat{V}_{PP}$ is the pseudopotential projector, and $\epsilon^{\sigma}_{n{\bf k}}$
are the KS eigenvalues.

The recalculation of $V_{HXC}^{\sigma}[n_i^{\sigma}]$ at each iteration of the self-consistency
for each atom and angular component would result in a major increase of computing cost. 
A large saving of time can be achieved by assuming a linear dependence of the SI potential 
on the occupation numbers:

\begin{eqnarray}
V_{HXC}^{\sigma}[n_i^{\sigma}]\>=\>p_i^{\sigma}\>
V_{HXC}^{\sigma}[n_i^{\sigma};p_i^{\sigma}=1]
\label{linear}
\end{eqnarray}

so that $V_{HXC}^{\sigma}[n_i^{\sigma};1]$ (i.e. the SI potential for the fully-occupied orbital) 
is set in the inizialization, and only the $p_i^{\sigma}$ needs to be updated during the 
self-consistency. Eq.\ref{linear} is exact for the Hartree term, which is the dominant contribution
for large occupation numbers, whereas it introduces a non-linearity error of 
${\cal O}(p_i^{1/3}-p_i)$ in the exchange-correlation part.\cite{pz} 

The time saving provided by the linear scaling argument is instrumental 
for calculating structural relaxations and electronic properties of large-sized 
systems. Furthermore, the assumption of linear scaling allows us to introduce relaxation effects 
in a very simple way into the pseudo-SIC scheme. Indeed if an electron state is localized 
its energy will change with the orbital occupation. Thus, in order to compare the calculated 
eigenvalue with the photoemission spectroscopy data (i.e. with the electron removal energy) the 
effects of the electron relaxation must be subtracted out of the one-electron potential. 
In DFT the energy required to remove a fraction $p$ of an electron from a one-electron localized 
state is:\cite{janak,pn,io} 

\begin{eqnarray}
\Delta E(p)\, =\, E(p) - E(0)\,=\,\int_{t=0}^{t=p}\,dt\,\epsilon(t),
\label{deltae}
\end{eqnarray}

where $\epsilon$ is the corresponding KS eigenvalue. 
The leading dependence of the LSDA KS potential (and eigenvalues) on the orbital occupations is 
indeed contained in its SI part. Thus, if $\delta \epsilon$ is the SI part of the LDA eigenvalue, 
within linear scaling we have $\Delta$E(p) = 1/2 $p^2$ $\delta\epsilon$(1). It follows that the SIC 
relaxation energy, 1/2 $p^2$ $\delta \epsilon$(1), can be subtracted out by rescaling the SIC potential 
as follows:

\begin{eqnarray}
V_{HXC}^{\sigma}[n_i^{\sigma}] \rightarrow {1\over 2}\>V_{HXC}^{\sigma}[n_i^{\sigma}].
\label{relax}
\end{eqnarray}

Through Eq.\ref{relax} we directly incorporate in the pseudo-SIC KS Equations the
electron removal energy due to the SI contribution. We point out that the eigenvalue relaxation is 
very important in order to match the photoemission spectroscopy results. Discarding this contribution, 
the SIC eigenvalues would strongly overestimate the electron removal energies. 

Notice that, if the KS eigenvalues did not depend on the occupation numbers, according to 
Eq.\ref{deltae} they would be equal to the electron removal energies. This is in fact the case
in Hartree-Fock calculations and in any other theory which obeys Koopman's theorem.
This property does not hold in LDA or any LDA-related scheme (such as GGA, SIC, etc.) where 
the potential explicitly depends on the orbital occupation numbers.

As is customary in atomic calculations, we assume the radial approximation for the atomic
orbital charges, so that the SIC projectors can be written:

\begin{eqnarray}
\gamma_{l,m_l,\nu}^{\sigma}({\bf r})\>= \>{1\over 2}\>p_{l,m_l,\nu}^{\sigma}\> 
V_{HXC}^{\sigma}[n_{l,\nu}^{\sigma}(r);1]\>
\phi_{l,m_l,\nu}^{\sigma}({\bf r}),
\label{rad}
\end{eqnarray}

where $\nu$ labels the atom type, and $n_{l,\nu}^{\sigma}(r)$ is the radial pseudo-charge 
density of orbital ($l$,$m_l$) (in radial approximation this does not depend on m$_l$).
Finally, the normalization coefficients are:

\begin{eqnarray}
C_{l,m_l,\nu}=\>{1\over 2}\>p_{l,m_l,\nu}^{\sigma}\>\int d{\bf r}\> 
V_{HXC}^{\sigma}[n_{l,\nu}^{\sigma}(r);1]\>
\left( \phi_{l,m_l,\nu}^{\sigma}({\bf r})\right)^2.
\label{rad1}
\end{eqnarray}

In summary, except for the electron occupation numbers, all the other atomic-like
ingredients can be imported from the code used for the pseudopotential generation,
and the Kleinman-Bylander projectors for the SIC are set during the initialization 
process. Since the calculation of the $p_{l,m_l,\nu}^{\sigma}$ through 
Eq.\ref{occ} is not very demanding, the global computational cost of the
pseudo-SIC for each self-consistent iteration is roughly equal to that of the 
basic LSDA. However within pseudo-SIC the number of self-consistent iterations required 
to reach the self-consistency can be larger than in LSDA due to oscillations 
in the values of the occupation numbers.

\subsection{Total Energy within pseudo-SIC}
\label{method2}

In the previous section we estabilished the form of the pseudo-SIC KS Equations.
Here we formulate a suitable expression for the total energy functional. We point 
out that, within our scheme, a physically meaningful energy functional which is
also related to Eqs. \ref{ks-sic} by a variational principle is not available. 
(It is, by construction, within LSDA and SIC-LSDA.) 


In SIC-LSDA the energy functional is:\cite{pz}

\begin{eqnarray}
E_{SIC}[n,m]\, =\, E[n,m]-\sum_{i,\sigma}\,E_{HXC}[n_i^{\sigma}]
\label{te_1}
\end{eqnarray}

where $E[n,m]$ is the LSDA energy functional and $E_{HXC}[n_i^{\sigma}]$ the Hartree 
exchange-correlation energy of the $i^{th}$ fully spin-polarized electron charge:

\startlongequation

\begin{eqnarray}
E_{HXC}[n_i^{\sigma}]\,=\,
\int d{\bf r}\> n_i({\bf r})\,\left({1\over 2}\,V_H[n_i^{\sigma}({\bf r})]+
{\cal E}_{XC}[n_i^{\sigma}({\bf r})]\right)\>,
\label{si_en}
\end{eqnarray}

where $V_H$ is the Hartree potential and ${\cal E}_{XC}$ is the local exchange-correlation energy density.
For the pseudo-SIC total energy we adopt the same expression as Eqs.\ref{te_1} and \ref{si_en}, with the 
orbital charges $n_i^{\sigma}$ given by Eqs.\ref{n_i} and \ref{occ}.

Notice that,

\begin{eqnarray}
{\delta E_{HXC}[n_i^{\sigma}]\over \delta p_i^{\sigma}}\,=\,C_i^{\sigma}.
\label{delta3}
\end{eqnarray}

Eq.\ref{delta3} represents the Janak thoerem\cite{janak,io} applied to the SIC contribution
since $C_i^{\sigma}$ (see Eq.\ref{c_i}) is the SI part of the atomic eigenvalue.
In terms of the pseudo-SIC eigenvalues (Eq. \ref{ks-sic}) the total energy can be rewritten:

\begin{eqnarray}
E_{SIC}[n,m]\,=\,\sum_{i,\sigma}\,f^{\sigma}_{n{\bf k}}\,\epsilon^{\sigma}_{n{\bf k}}
-\sum_{\sigma}\,\int \! d{\bf r}\> n^{\sigma}({\bf r})\,V^{\sigma}_{HXC}[n({\bf r}),m({\bf r})]
+E_{HXC}[n,m]+E_{ion}
\nonumber
\end{eqnarray}
\begin{eqnarray}
+\,{1\over 2}\sum_{n{\bf k},\sigma}\>f^{\sigma}_{n{\bf k}}\,
\langle\psi_{n{\bf k}}^{\sigma}\,|\,\hat{V}^{\sigma}_{SIC}\,
|\,\psi_{n{\bf k}}^{\sigma}\rangle
-\,\sum_{i,\sigma}\,E_{HXC}[n_i^{\sigma}],
\label{te_2}
\end{eqnarray}

\stoplongequation

where $E_{ion}$ is the usual Ewald term. $\sum_{i,\sigma}E_{HXC}[n_i^{\sigma}]$ produces a gentle 
modification 
to the LSDA energy functional whereas the fifth term in Eq.\ref{te_2} is a strongly varying contribution 
which compensates the same contribution present in the pseudo-SIC eigenvalues. 
Without this compensation, Eq.\ref{te_2} would give very inaccurate total energies which would be 
unphysically far from the LSDA values. 
A numerical example of this behavior will be shown in Section \ref{res1}.

\section{Results}
\label{results}

\subsection{Technical details}
\label{technic}

In this work we compare results from our plane wave USPP\cite{van} pseudo-SIC 
implementation with results from conventional LSDA USPP method. 
The local exchange-correlation energy functional is modeled using the the 
Perdew-Zunger interpolation formula.\cite{pz} The use of USPP\cite{van} allows us 
to obtain well-converged results with moderate cut-off 
energies (35 Ry for MnO and YMnO$_3$, 40 Ry for ZnO and 45 Ry for GaN).
In order to have highly transferable USPP, two projectors per angular channel
are included for all the atoms. The 3d$^{10}$ electrons in ZnO and GaN are treated as valence 
states, while the semicore Y $s$ and $p$ electrons are placed in the valence for the LSDA
calculations and in the core for the pseudo-SIC calculations (the reason will be explained 
in the following section).  
For total energy calculations we use up to 8$\times$8$\times$8 grids of special k-points.\cite{mopa}

\subsection{Atomic ingredients for the pseudo-SIC}
\label{atomic}

The atomic quantities necessary to build the pseudo-SIC hamiltonian are the pseudo atomic
orbitals $\phi_{l,m_l,\nu}$, required in the Kleinman-Bylander projector and for the 
calculation of $p_i^{\sigma}$, and the SIC potentials $V_{HXC}^{\sigma}[n_{l,\nu}^{\sigma}(r);1]$ 
for the respective pseudo orbital charges calculated at full electron occupation (we
use pseudopotential and not all-electron atomic functions for obvious reasons of smoothness.)

\begin{figure}
\narrowtext
\epsfxsize=8.5cm
\centerline{\epsffile{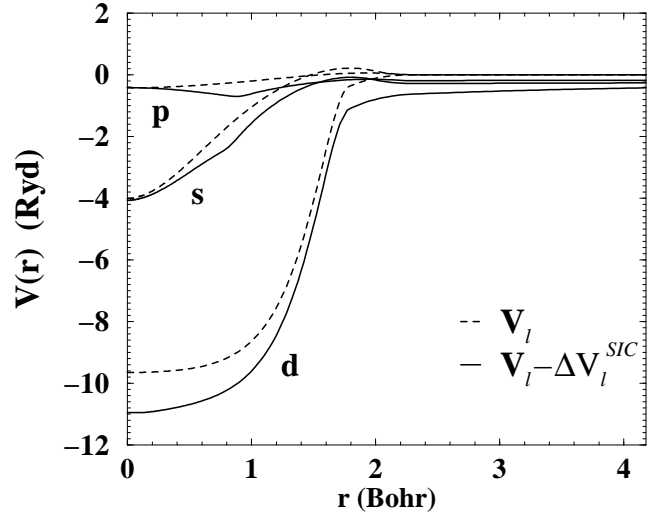}}
\caption{Zn pseudopotentials (V$_l$) and the corresponding atomic SIC ($\Delta V_l^{SIC}$) 
calculated at full orbital occupation. The correction lowers the one-electron potential
and recovers the physical long-range limit -2/r which is missing in LDA.
\label{zn}}
\end{figure}

To illustrate the impact of the atomic SIC, in Figure \ref{zn} we compare the Zn s, p, and d 
(unpolarized) pseudopotentials (V$_l$) to the same quantities minus the 
corresponding corrections $\Delta V_l^{SIC}=V_{HXC}[n_{l,\nu};1]$. As expected, the SIC makes 
the electron potentials more attractive and recovers the physical long-range limit -2/r (in Ry). 

\begin{figure}
\epsfxsize=9.0cm
\centerline{\epsffile{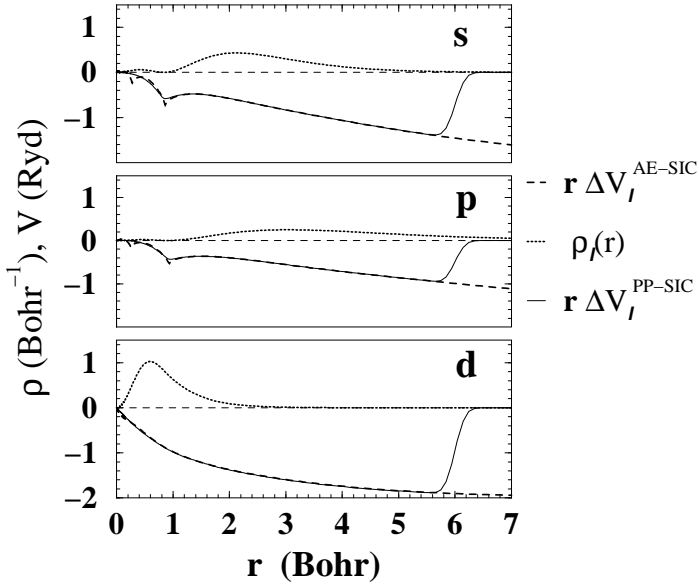}}
\caption{Zn atomic SIC for pseudo (solid line) and all-electron (dashed line) orbital
charges. The pseudo orbital charges (dotted lines) are plotted on the positive y-axis. 
The pseudo SIC are cut-off at a diagnostic radius where the orbital charges vanish.  
\label{zn.sic}}
\end{figure}

Note that, before transfering $V_{HXC}[n_{l,\nu};1]$ to the crystal, the long-range tails 
must be cut off, since the SIC should act as a local correction: each SIC potential 
must be directly applied only to the electron states localized on the same atom, otherwise
the overlapping Coulomb tails would give rise to an unphysical SIC overestimation.
In Figure \ref{zn.sic} we report $V_{HXC}[n_{l,\nu};1]$ (times $r$) for both pseudo and all-electron 
orbital charges (the latter show the cusps corresponding to the wavefunction nodes),
and the pseudo orbital charges $\rho_l$. Since only the products $V_{HXC}[n_{l,\nu}]\,\rho_l$ contribute
in the Kleinmann-Bylander construction, the SIC potentials can be cut off as soon as 
the orbital charges vanish, without losing accuracy. 

It is important to notice that $V_{HXC}^{\sigma}[n_{l,\nu}^{\sigma};1]$ is not very 
sensitive to the atomic valence configuration from which it is actually generated. 
For example the integrated value of $V_{HXC}^{\sigma}[n_{3d};1]\cdot\phi^2_{3d}$ 
(i.e. the atomic SI) changes by only $\sim$ 1\% if calculated in the 3d$^{10}$4s$^2$, 
the 3d$^{10}$4s$^1$, or the 3d$^{10}$4s$^0$ configuration. In other words, despite being 
valence dependent, $V_{HXC}^{\sigma}[n_{l,\nu};1]$ is fairly transferable. 

Notice also that the pseudo-SIC does not interfere with the pseudopotential construction, 
and the same pseudopotentials used for the LSDA calculations can be used for the pseudo-SIC 
as well, due to the presence of the occupation numbers. Indeed, in the limit $p_i$=0 
the SIC vanishes, and the hamiltonian must continuously recover its LSDA value.

However for pseudo-SIC calculations we prefer to build pseudopotentials which take 
into account the SIC for the core states. Although this correction does not significantly modify 
the valence properties,\cite{note} there are reasons related to pseudopotential 
transferability which suggest that the SIC should be included in the core states. 
First, the SIC could push strongly localized valence states (e.g. the semicore states) down in 
energy and unphysically close to the core states if the latter are not shifted down consistently. 
Second, the application of SIC to the semicore states may better justify their inclusion in the 
core even if in LSDA they need to be treated as valence states. This is the case for the Y 4s 
and 4p electrons, and the Ga 3d electrons. 

A third reason is even more fundamental: increasing the binding energy and the space 
localization of the core states makes them less sensitive to charge relaxation or chemical 
activity. This reinforces the hypothesis that core states are insensitive to the chemical 
environment, which is at the basis of the pseudopotential approximation.
  
The transferability of the atomic SIC to the extended system can be tested with the same 
procedure used for the pseudopotentials: the atomic pseudo-SIC hamiltonian should 
recover the all-electron, fully self-consistent SIC-LSDA eigenvalues. We generally find
that pseudo-SIC and SIC-LSDA atomic eigenvalues agree within $\sim$ 1\%

Finally we stress a fundamental point: our procedure is not in any way restricted to the use 
of pseudopotentials or any other particular choice of basis functions. For example it can be 
equally well implemented within the context of all-electron methods. The advantage of using the 
pseudopotential approach relies on the fact that the SIC projectors can be easily assembed 
from quantities (such as the projections of Bloch states onto atomic orbitals) which are usually 
calculated in any pseudopotential-based computing code.
 

\subsection{Wide-gap semiconductors}
\label{res1}

We begin our series of pseudo-SIC applications with two of the most prototypical wide-gap
semiconductors, wurtzite ZnO and GaN. These materials are ideally suited as test cases
since the LDA predictions for the band energies show some easily recognizable discrepancies 
with the abundantly available photoemission spectroscopy data. Furthermore, these
compounds are of technological interest due to their applications in optoelectronic and 
piezoelectric devices. Nowadays the electric polarization and related properties can be efficiently 
calculated by first-principles techniques.\cite{polar} Thus a theory able to repair the LDA 
description of the electronic structure will also be important for the accurate evaluation 
of dielectric and piezoelectric response.

Since ZnO and GaN show similar structural and electronic behavior, we describe them 
together. In Table \ref{tab} we report our calculations within LDA and pseudo-SIC for the 
equilibrium structural parameters. These have been evaluated by energy minimization within 
the full space of parameters a, c and u (u is the anion-cation distance in units of c). 
In general, the SIC volume is larger than the LDA volume, since the stronger electron 
localization due to the SIC enhances the electron screening and reduces the electron-ion 
interaction and the charge hybridization. This is generally a favorable correction since it 
is well known that within LDA (or LSDA) the lattice parameters are underestimated compared with 
the experimental values. 

\begin{table}
\caption{Equilibrium lattice parameters (a,c,u), band gap (E$_g$) and average d-band energy (E$_d$) 
for ZnO and GaN. All the distances are in Bohr except the anion-cation distance, u, which is in units 
of c. All the energies are in eV. For ZnO the experimental values
are from Ref. {\protect\onlinecite{book}}.
\label{tab}}
\centering\begin{tabular}{ccccc}
                     & LDA & pseudo-SIC & Expt. \\
\hline\hline
   ZnO         & \multicolumn{3}{c} {} \\
\hline
   a            & 6.12        & 6.17  & 6.16     \\
   c            & 9.88        & 9.79  & 9.84   \\
   u            & 0.378       & 0.384 &  0.382   \\
   E$_g$        & 0.94        & 3.70  &  3.4     \\
   E$_d$        & --5.3       & --7.5 & -7.8     \\
\hline
   GaN               & \multicolumn{3}{c} {}  \\
\hline
   a            & 6.03   & 6.045 & 6.03{\protect\cite{lamb}}   \\
   c            & 9.80   & 9.81  & 9.80{\protect\cite{lamb}}    \\
   u            & 0.377  & 0.378 & 0.375{\protect\cite{lamb}} \\
   E$_g$        & 2.16   & 4.26  & 3.5{\protect\cite{book}}     \\
   E$_d$        & --13.8 & --18.1 & --17.1{\protect\cite{lssmamr}} \\
\end{tabular}
\end{table}

In the case of ZnO and GaN, however, our LDA values are already in excellent agreement 
(within less than $\sim$ 1\%) with the experiments and no correction would be required. 
Happily, the pseudo-SIC calculations do not overcorrect the LDA values.
(the reason will be explained later on the basis of the band structure results).
This is an important aspect since typical beyond-LDA methodologies, also not derived
by variational principles, often improve the LDA description of the electron 
excitation spectra but are less accurate than LDA in the determination of the 
equilibrium structure.

\begin{figure}
\epsfxsize=8.5cm
\centerline{\epsffile{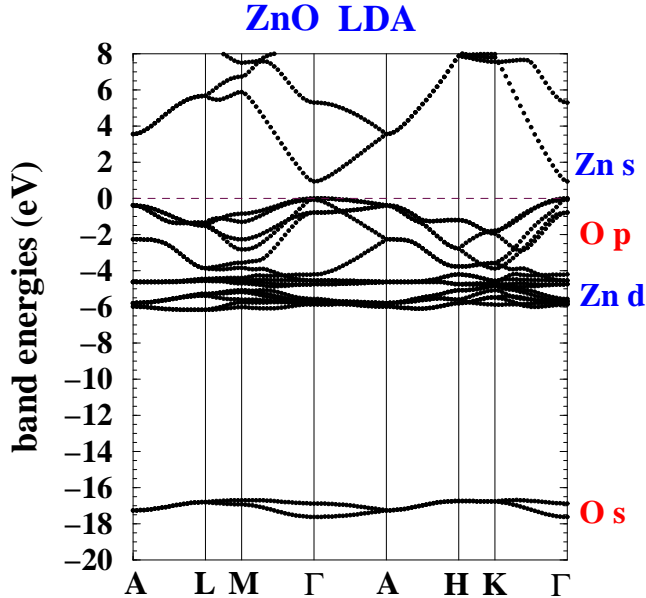}}
\caption{LDA band structure for wurtzite ZnO. The orbital character for each group
of bands is indicated.
\label{zno}}
\end{figure}

In Figure \ref{zno} our calculated LDA energy band structure of wurtzite ZnO is shown. 
For such strongly ionic compounds each group of bands can be clearly labeled according to a single, 
dominant orbital character as indicated in the Figure.

With respect to the photoemission spectroscopy results, the 3d bands calculated within LDA are too high in energy 
and overlap with the sp$^3$ valence band manifold. This produces a spurious p-d hybridization which shrinks
the energy range of the sp$^3$ bands. Furthermore, there is the notorious problem of the fundamental energy gap 
which is underestimated by $\sim$ 40 \% (see Table \ref{zno}).

\begin{figure}
\epsfxsize=8.5cm
\centerline{\epsffile{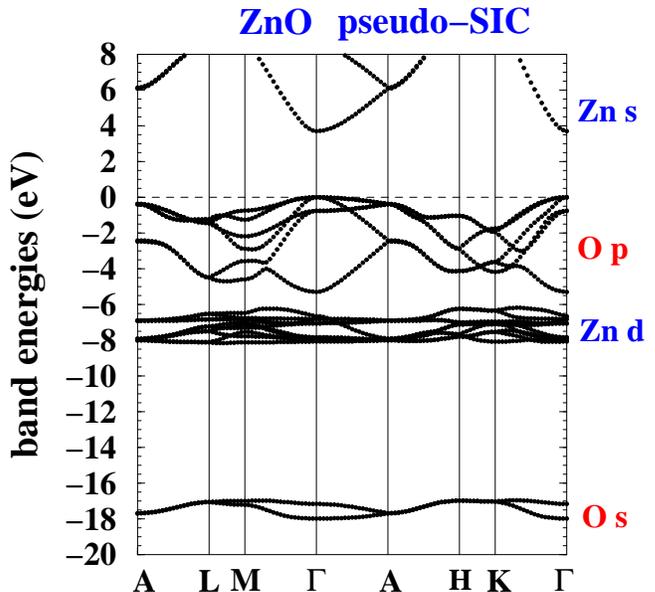}}
\caption{Pseudo-SIC band structure for wurtzite ZnO
\label{zno.sic}}
\end{figure}

All these undesirable features are largely overcome in our calculated pseudo-SIC band structure 
(Fig. \ref{zno.sic}). The largest SIC effect is a downshift of the d band energies, which are now placed 
$\sim$ 3 eV below the center of the sp$^3$-band manifold, in excellent agreement with the experiments. 
Accordingly, the sp$^3$ band manifold is $\sim$ 1 eV broader than in LDA.

Furthermore, the bands with O p orbital character, which are the main contributors to the valence band top (VBT),
are also corrected for an amount of SI corresponding to their average electron occupation 
($\sim$ 0.8 for the O p orbitals). In contrast the conduction band bottom (CBB), which is mainly Zn s in character, 
is almost unchanged, since the Zn s orbital occupation is $\sim$ 0.1. As a consequence the pseudo-SIC energy gap 
opens up to a value fairly close to the experimental band gap.

Our results show a substantive agreement with the calculated SIC ZnO band structure of Ref.\onlinecite{vkp}
(they obtain E$_g$ = 3.5 eV). However the two methods act in quite different ways: in the approach of 
Ref.\onlinecite{vkp} all the bands 
(occupied and unoccupied) are corrected by the atomic SIC potential of the corresponding fully occupied 
orbital charges. Thus in Ref.\onlinecite{vkp} the increase of the energy gap with respect to the LDA can 
be roughly quantified as the difference between the SIC of the fully occupied Zn 4s and O 2p atomic eigenvalues. 
Instead, in our scheme only the occupied bands are corrected. We belive that this is a more conceptually 
sound approach for the reasons explained in Section \ref{method1}.
 
\begin{figure}
\epsfxsize=8.5cm
\centerline{\epsffile{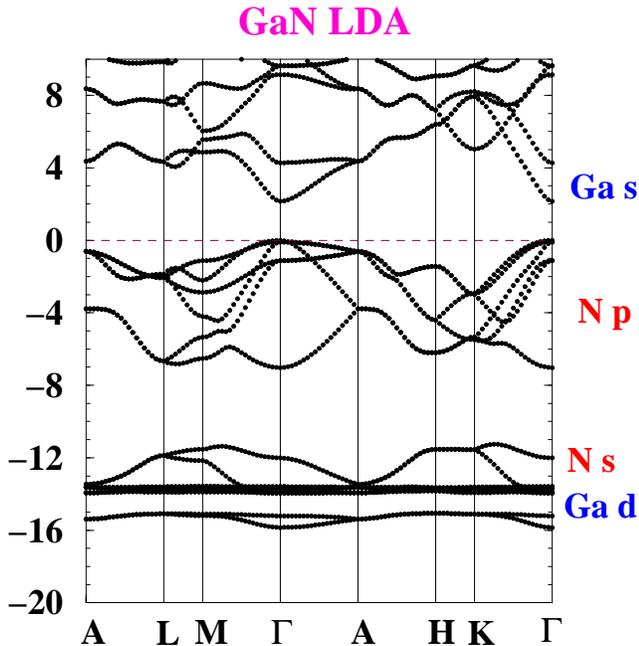}}
\caption{Band structure of wurtzite GaN calculated in LDA.
\label{gan}}
\end{figure}

In Figure \ref{gan} we show the LDA band structure of wurtzite GaN. As is usual in LDA calculations, the Ga d 
band energies fall within the same energy range as the N s bands, in disagreement with the experiments which place the d 
bands $\sim$ 3 eV below the N s bands.\cite{lssmamr} This gives rise to a spurious s-d hybridization which 
causes an increase of lattice constant due to the closed-shell repulsion enhanced by the resonances between 
states on neighboring sites\cite{fms,wn,cheli} (i.e. N s and Ga d). In our calculation this effect compensates 
the LDA tendency of underestimating the lattice parameters, and causes a better-than-usual agreement with the 
experimental data (Tab. \ref{tab}). However, the description of the band structure appears grossly inappropriate 
when compared to the X-ray photoelectron spectra.\cite{lssmamr} The s-d hybridization splits the N s bands into two 
sections, one placed above and the other below the d bands. The s-d manifold is positioned only $\sim$ 6 eV below 
the bottom of the valence sp$^3$ manifold, and the energy gap is 40\% smaller than the experimental value. 

\begin{figure}
\epsfxsize=8.5cm
\centerline{\epsffile{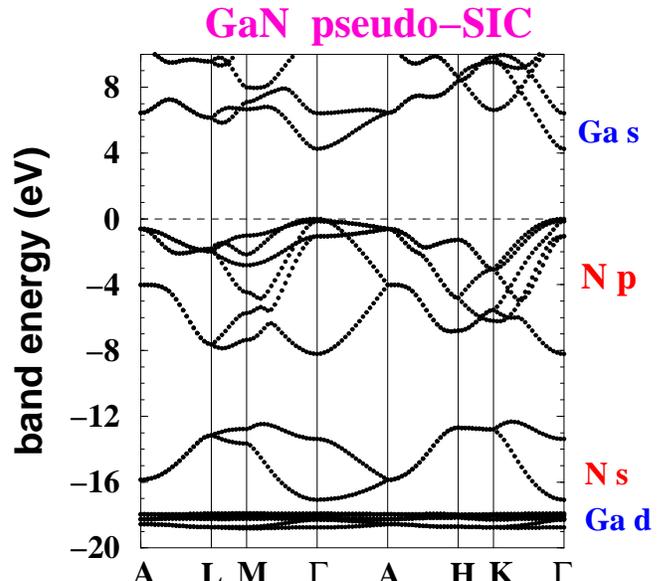}}
\caption{Band structure of wurtzite GaN calculated in pseudo-SIC.
\label{gan.sic}}
\end{figure}

The pseudo-SIC corrects in large part the LDA description. In Figure \ref{gan.sic} we show the GaN band structure 
calculated within pseudo-SIC. Here the Ga d bands (E$_d$ $\sim$ -18 eV) are correctly located $\sim$ 3 eV below the 
center of the N 2s bands, thus the spurious s-d hybridization is avoided and the closed-shell repulsion reduced. 
This explains why the equilibrium lattice parameters calculated within pseudo-SIC are close to the LDA values 
although the pseudo-SIC is generally expected to enhance the electron-ion screening.

Also, the binding energies of Ga d and N s states are increased by $\sim$ 4 eV and 2 eV respectively, thus
correcting in large part the faults of the LDA description. However within pseudo-SIC the energy gap is somewhat 
overcorrected (it is $\sim$ 0.9 eV larger than the experimental value). 

\begin{figure}
\epsfxsize=6cm
\centerline{\epsffile{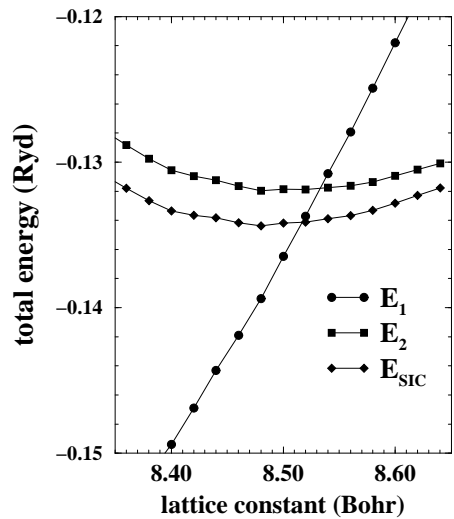}}
\caption{Total energy values for zincblende GaN as a function of the lattice constant a$_0$. The
meaning of the curves is explained in the text.
\label{gan.en}}
\end{figure}

To conclude this Section, in Figure \ref{gan.en} we analyze the behavior of the pseudo-SIC total energy 
(Eq.\ref{te_2}) for zincblende GaN as a function of the lattice constant in the region around 
its experimental 
value a$_0$=8.505 Bohr. E$_1$ is the functional given by the first four terms of Eq.\ref{te_2}. 
Clearly, its minimum value is far out of the physically meaningful region. E$_2$ is obtained by adding 
to E$_1$ the fifth term of Eq.\ref{te_2}, thus E$_2$ equals the LSDA energy functional (except for 
the difference in the wavefunctions). This functional has a minimum value in excellent agreement with 
the experiments. Finally, E$_{SIC}$ is 
E$_2$ + $\sum_{i,\sigma}$ E$_{HXC}[n_i^{\sigma}]$. This last term is almost independent of the lattice 
parameter, since it only changes through the $p_i^{\sigma}$, which are rather local quantities. 
As a consequence, E$_{SIC}$ and E$_2$ have similar behavior.


\subsection{Transition-metal oxides.}
\label{res2}

The LSDA misrepresentations of the electronic properties of the transition-metal oxides 
originate mainly from the description of the d electron states. At variance with the non-magnetic 
semiconductors considered in the previous section, in the transition-metal oxides the d states lie 
higher in energy and closer to the fundamental gap, often overlapping with the oxygen p states. 
Since in LSDA the d electron binding energies are severely underestimated, the d character is generally 
dominant at the VBT, so these compounds are described as Mott-Hubbard insulators. 
This is in striking contrast with the photoemission spectroscopy data which describe these materials as 
charge-transfer insulators (or eventually in the intermediate charge-transfer Mott-Hubbard regime) with 
a majority or even dominant O p character at the VBT.\cite{zscrhbd}

Furthermore the LSDA energy gaps for these materials are even more severely underestimated than 
those of the III-V or II-VI semiconductors and in some cases the gap can be vanishing. 
This is because the major driving force leading to the formation of a gap separating bands of equal 
orbital character should be the on-site Coulomb energy U, but in LSDA the energy gap can only open 
due to Hund's rule and the crystal field splitting. These are both of order 1 eV,  i.e. almost one order 
of magnitude smaller than U, and therefore can be easily overcome by the the band dispersion. 

Finally, since the LSDA tends to broaden the space distribution of the electron charge 
and emphasize the interatomic charge hybridizations, the local magnetic moments are 
generally underestimated with respect to the experimental values.

\begin{figure}
\epsfxsize=8.5cm
\centerline{\epsffile{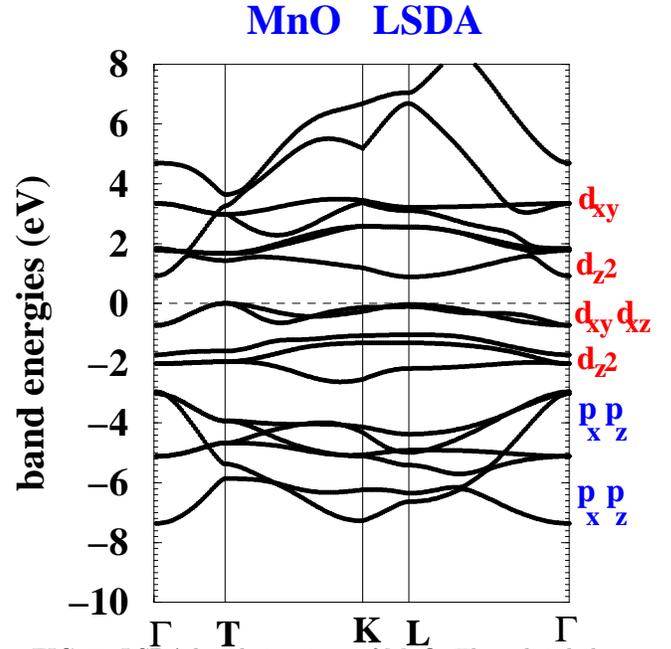}}
\caption{LSDA band structure of MnO. The orbital character is indicated for each band. The orbitals
are expressed in hexagonal coordinates, i.e. $z$ is parallel to the [111] direction.
\label{mno}}
\end{figure}

As applications of the pseudo-SIC to magnetic compounds we choose two of the most widely studied 
transition-metal monoxides, MnO and NiO. For these materials we can compare our results 
with both experimental data and the output of some common methodologies used for repairing 
the faults of the LSDA description.\cite{towk,zscrhbd} These schemes include the 
SIC-LSDA,\cite{sg,ls,af} the LDA+U,\cite{aza,ls,st,slp,babh} and the model GW.\cite{gb,mcpb,mpbr,cmp} 

The ground state of MnO and NiO is characterized by A-type antiferromagnetic ordering, which consists 
of (111) ferromagnetic layers of Mn (or Ni) with alternating spin directions, while the oxygens are non-magnetic. 
The symmetry is rhombohedral, with 4 atoms per primitive unit cell. In our calculations we fix the lattice constant 
to its experimental value (a$_0$ = 8.37 Bohr and 7.93 Bohr for MnO and NiO, respectively).

In Figure \ref{mno} the LSDA band structure for MnO is shown. Notice that the orbital characters are referred 
not to the cubic, but to the hexagonal coordinates, i.e. $x$ and $y$ lie on the hexagonal (111) plane, 
and $z$ is parallel to the [111] axis. Also, within the hexagonal (or rhombohedral) crystal field, the 3d 
orbitals are split into two doublets (d$_{xy}$, d$_{x^2-y^2}$) and (d$_{xz}$, d$_{yz}$), and one singlet (d$_{z^2}$). 

The nominal ionic configuration Mn$^{2+}$O$^{2-}$ suggests a Hund's rule-induced splitting between
filled spin up and empty spin down 3d bands, resulting in an energy gap.
From the analysis of the band character we see that the VBT is composed of 60\% (d$_{xy}^{\uparrow}$, 
d$_{x^2-y^2}^{\uparrow}$) and (d$_{xz}^{\uparrow}$, d$_{yz}^{\uparrow}$) orbitals, with the 
remaining 40\% coming evenly from the (p$_x$, p$_y$) doublets of the two equivalent oxygens. 
In contrast, the conduction band bottom (CBB) consists of $\sim$ 80\% d$_{z^2}^{\downarrow}$ centered 
on the same Mn, which is equivalent to the d$_{z^2}^{\uparrow}$ orbital localized on the other Mn. 
Thus, from the LSDA calculation, a picture of the MnO as a small-gap Mott-Hubbard insulator emerges, which
is not what should be expected according to the experiments. 

Furthermore, while the value of the Mn magnetic moments is satisfactory, the size of the LSDA energy gap severely 
underestimates the measured value (see Table \ref{tab2}). Also, notice that the occupied 3d bands are 
rather flat and well separated by a $\sim$ 1 eV gap from the underlying O 2p band manifold which is $\sim$ 6 eV wide. 

\begin{figure}
\epsfxsize=8.5cm
\centerline{\epsffile{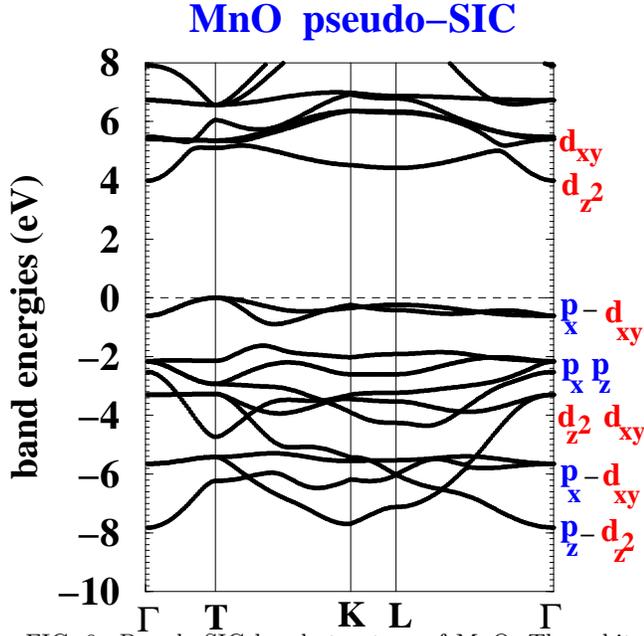}}
\caption{Pseudo-SIC band structure of MnO. The orbital character is indicated for each band.
\label{mno.sic}}
\end{figure}

In Figure \ref{mno.sic} we see how the application of the pseudo-SIC modifies the results. First of all,
there is no longer a gap between 3d and 2p bands. The SIC causes a downward shift of the Mn 3d bands with
respect to the more dispersed O 2p bands, thus the 3d character is now spread across the whole (8 eV wide) 
valence band manifold, and is heavily mixed with the O 2p character. 

This shift significantly increases the energy gap which is now well 
within the experimental uncertainty (see Table \ref{tab2}). It also changes the VBT character, which 
is now composed of 40\% 3d doublets and 60\% 2p orbitals. 85\% of the CBB still comes from the 
d$_{z^2}^{\downarrow}$ orbital. 

Thus the agreement with experiments, which locate the MnO in the intermediate charge-transfer Mott-Hubbard regime, 
is restored. Furthermore, the enhanced on-site localization of the 3d orbitals due to the SIC increases the magnetic 
moment which is now in excellent agreement with the experimental value.

\begin{table}
\caption{Magnetic moments, M, and energy band gaps, E$_g$, of MnO and NiO. The upper part shows our results 
calculated within LSDA and pseudo-SIC, in comparison with the experimental values.
The lower part shows results of other beyond-LSDA calculations (results in parentheses are explained in the
text).
\label{tab2}}
\centering\begin{tabular}{ccccc}
                       & \multicolumn{2}{c} {MnO}  & \multicolumn{2}{c} {NiO} \\
                       &   E$_g$(eV) &  M($\mu_B$) & E$_g$(eV) &  M($\mu_B$) \\   
\hline
   LSDA                 &      0.92   &       4.42  &      0.4   &    1.11     \\
  pseudo-SIC            &      3.98   &  4,71       &   3.89     &    1.77      \\
  Expt.             &  3.8-4.2{\protect\cite{vanelp}} &  4.79,{\protect\cite{fjw}} 4.58{\protect\cite{ch}} 
         & 4.0,{\protect\cite{horh}} 4.3{\protect\cite{fm}} &  1.77{\protect\cite{fjw}} 1.90{\protect\cite{ch}} \\
\hline\hline
  SIC-LSDA\cite{sg,stw} &      3.98   &    4.49     &   2.54     &    1.53      \\
  SIC-LSDA\cite{af}     &  6.5(3.4)   &    4.7(4.7) &  5.6(2.8)  & 1.7(1.5)    \\
  LDA+U\cite{aza}       &  3.5        &    4.61     &  3.1       &  1.59      \\
  LDA+U\cite{babh}      &             &             &  4.1(2.8)  & 1.83(1.73)     \\
  LDA+U\cite{slp}       &             &             &    3.38    &      1.69        \\
  GW\cite{mcpb}       &      4.2   &   4.52      &   3.7        &   1.56    \\
\end{tabular}
\end{table}

The LSDA description of NiO is even worse than that of MnO. Indeed, the Ni$^{2+}$ ion has nominally 
eight electrons in the occupied bands, thus the energy gap within the d band manifold can only be opened 
by the weak crystal field splitting.

\begin{figure}
\epsfxsize=8.5cm
\centerline{\epsffile{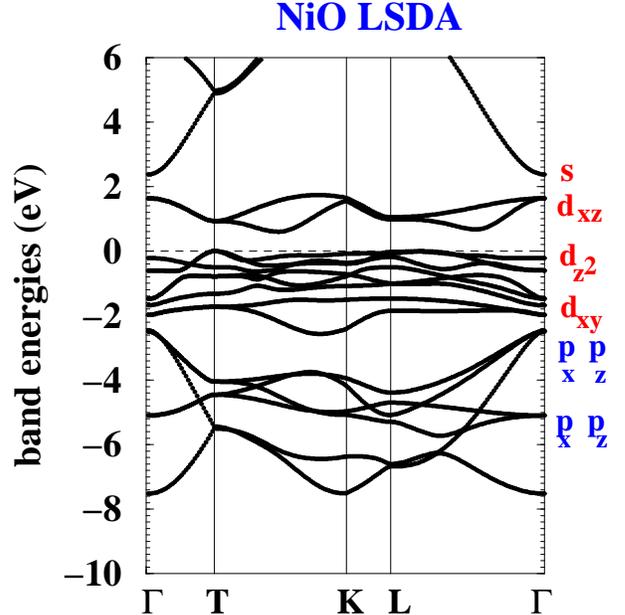}}
\caption{LSDA band structure of NiO. The orbital character is indicated for each band.
\label{nio}}
\end{figure}

In Figure \ref{nio} we show the band structure of NiO obtained within LSDA.
The fundamental gap occurs between bands derived from the singlet d$_{z^2}^{\uparrow}$ and the doublet 
(d$_{xz}^{\downarrow}$,d$_{yz}^{\downarrow}$). In particular, the VBT is almost purely (90 \%) d$_{z^2}^{\uparrow}$ 
in character, and no hybridization with oxygens is present. The gap value (see Table \ref{tab2}) is only 0.4 eV,
and in fact some LSDA calculations even describe NiO as a metal.\cite{af} The majority d bands are rather flat and, 
as in MnO, separated by a $\sim$ 1 eV wide gap from the p bands. Finally, the magnetic moments are underestimated 
by $\sim$ 40\%. Thus, according to the LSDA calculations, NiO is a small-gap Mott-Hubbard insulator, whereas the 
experiments describe NiO as a wide-gap charge-transfer insulator.

\begin{figure}
\epsfxsize=8.5cm
\centerline{\epsffile{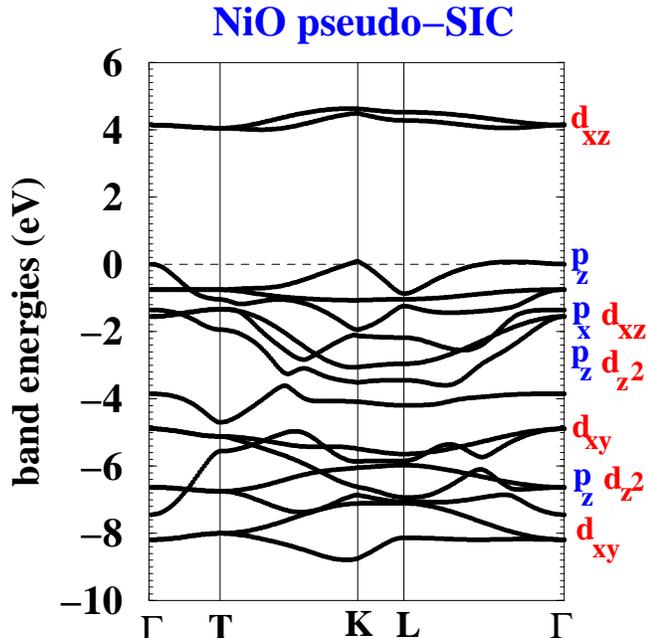}}
\caption{Pseudo-SIC band structure of NiO. The orbital character is indicated for each band.
\label{nio.sic}}
\end{figure}

The inclusion of SIC (Figure \ref{nio.sic}) greatly improves the NiO description. As in MnO, 
the eight occupied d bands are shifted down in energy and strongly hybridized with the O p bands. 
The VBT becomes a mixture of d and p character, with a predominance of the latter.   
For example at $\Gamma$ the VBT singlet is almost purely p$_z$, whereas at K the twofold 
degenerate VBT comes from the hybridization between (p$_x$, p$_y$), which contribute 80\%, 
and (d$_{xz}^{\uparrow}$, d$_{yz}^{\uparrow}$), which contribute 20\%. Thus, it is a pd$\pi$-hybridized 
band. The same pd$\pi$ hybridization characterizes the CBB doublet, but the percentages of the composition 
are reversed: 90\% of the charge comes from (d$_{xz}^{\downarrow}$, d$_{yz}^{\downarrow}$), and only 10\% 
from (p$_x$, p$_y$). Thus, according to the pseudo-SIC, NiO is a charge-transfer insulator. Furthermore, 
the calculated energy gap and magnetic moments are in good agreement with experiments (see Table \ref{tab2}).
As is evident from the band structure, the Ni magnetic moment, M=1.77 $\mu_B$, comes from the 
spin-polarization of the degenerate orbitals d$_{xz}$ and d$_{yz}$, each of them carrying M/2 magnetization.
Finally, the bands at the bottom of the occupied p-d manifold are 90\% from the doublet 
(d$_{xy}^{\uparrow}$, d$_{x^2-y^2}^{\uparrow}$). These orbitals lie on the hexagonal plane and
are entirely localized since they do not point towards the oxygens. As a result their corresponding 
bands experience the largest SIC. 

A number of beyond-LSDA approaches have been attempted in the past and tried on the series of 
transition-metal oxides. In Table \ref{tab2} we listed the results of some of these calculations. 

Whithin full SIC-LSDA, the results for the transition-metal oxides are critically sensitive to 
the choice of orbitals taken as localized. In Ref.\onlinecite{af} (a LMTO-ASA calculation) 
the SIC-LSDA band structures are calculated in two ways: The first calculation 
assumes that 3d and the 2p orbitals are both localized and therefore affected by the SI, 
and the second is performed with only the 3d orbitals taken as localized. (In the Table we
report the results of both the calculations; the d-only SIC-LSDA results are in parentheses).
It is shown that the first option gives a band structure in better agreement with photoemission 
spectroscopy, with the exception of the band gap which is strongly overestimated. Instead, the choice 
of correcting only the d orbitals gives a better energy gap but also leads to a too large 
downshift of the d bands with respect to the p bands (e.g. for MnO and NiO the d bands lie $\sim$ 5-6 eV 
below the center of the p bands\cite{af}). As a consequence, the calculation exaggerates the charge-transfer 
character of the band gap.

The strategy of correcting both d and p orbitals represents our preferred point of view,\cite{reason} 
and is consistent with the pseudo-SIC which assumes that all the electron states should be corrected, 
without any a-priori discrimination. Indeed, the pseudo-SIC band structures are similar to those  
obtained in Ref.\onlinecite{af} with the first choice of orbitals, with the important exception 
that in our pseudo-SIC the energy gap is in much better agreement with experiments. We speculate
that the reason for this difference is the absence in SIC-LSDA of the relaxation energy contribution, 
which shifts the occupied d and p bands up in energy.

The LDA+U corrects only the d states, and  it shows similarities with the d-only SIC-LSDA in 
describing the properties of the transition-metal oxides.\cite{aza,ls,st,babh} Indeed in both 
these approaches, the transition-metal oxides are all described as charge-transfer insulators.
However in LDA+U the amplitude of the d band shift is critically controlled by the parameter U.
For example, in Ref.\onlinecite{babh} (a projector augmented-wave (PAW) calculation) it is shown that 
for NiO the widely used value U=8 eV gives a good energy gap but a poor description of the occupied 
band manifold due to the exaggerated downward shift of the d bands. In contrast, the value U=5 eV gives a 
smaller energy gap, but optical properties and magnetic moments in better agreement with experiments.
(Results for U=5 eV are reported in parentheses in Table \ref{tab2}).

Finally, we report the results of the model GW,\cite{gb} an approach radically 
different from both the SIC or the LDA+U schemes, based on a model self-energy correction to the 
LSDA KS equations. The model GW is capable of giving an accurate description of both 
transition-metal oxides\cite{mcpb,mpbr} and other non-magnetic semiconductors,\cite{gb,mcpb,cmp} 
thus it should probably be considered as the most reliable reference. The fact that the pseudo-SIC values for 
the energy gaps are close to the results obtained within GW is an indication that, at least for this 
family of compounds, the SI, and not many-body effects, really represents the main source of the LSDA error. 

\subsection{Hexagonal manganites}
\label{res3}

As a final application of the pseudo-SIC, we consider hexagonal YMnO$_3$, which has 
recently attracted the attention of both the experimental\cite{hcs,katsu,fpp,flfgp} 
and the theoretical\cite{medved,fh} 
communities since it shows the uncommon characteristic of being both magnetically and ferroelectrically 
ordered within the same bulk phase. The study of YMnO$_3$ is motivated by the possible applications 
of ferroelectromagnetic materials as building blocks for spintronic devices\cite{spintr} and, from a more 
fundamental standpoint, by the necessity to understand the interaction between magnetic and ferroelectric polarization 
and the conditions which favor this coexistence.\cite{nicola,fh}

YMnO$_3$ can be grown in both orthorhombic\cite{iliev1} and hexagonal\cite{yakel,klk,aken} structures, 
although the latter is the most stable. The orthorhombic phase, typical of manganese perovskites (e.g. LaMnO$_3$),
is antiferromagnetic but not ferroelectric. The hexagonal phase shows a spontaneous ferroelectric 
polarization ${\bf P}$ $\sim$ 5.5 $\mu$C/cm$^2$ parallel to the c axis\cite{fiyi,klk} below a critical 
temperature T$_c$=900 K, and is A-type antiferromagnetic\cite{siko,fro} at temperatures below T$_N$=80 K. 
In this section we will consider only hexagonal YMnO$_3$.

\begin{figure}
\epsfxsize=4cm
\centerline{\epsffile{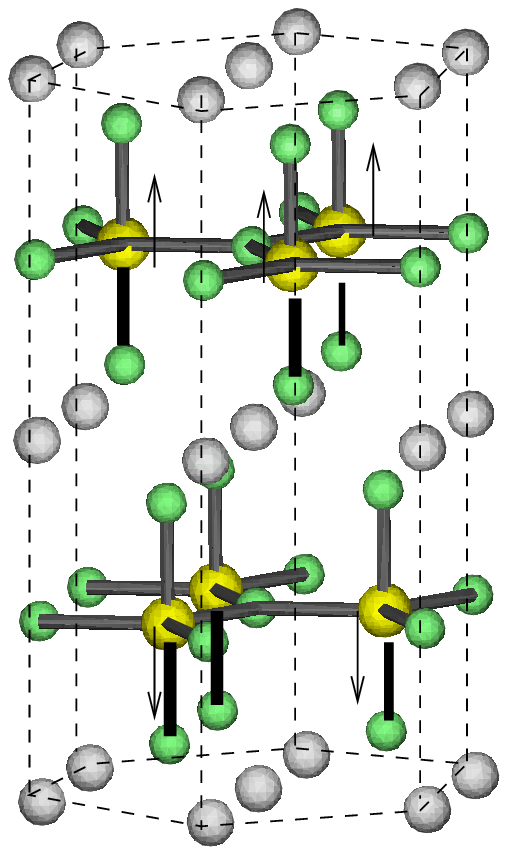}}
\caption{Structure of YMnO$_3$ in the paraelectric P6$_3$/mmc phase. The arrows are placed on the Mn ions 
and indicate the spin-polarization direction (thus the ordering is A-type antiferromagnetic). Each Mn is
surrounded by a bi-pyramidal cage of five corner-sharing oxygens. Atoms not connected by bonds are Y.
\label{ymn.fig}}
\end{figure}

\begin{figure}
\epsfxsize=9.0cm
\centerline{\epsffile{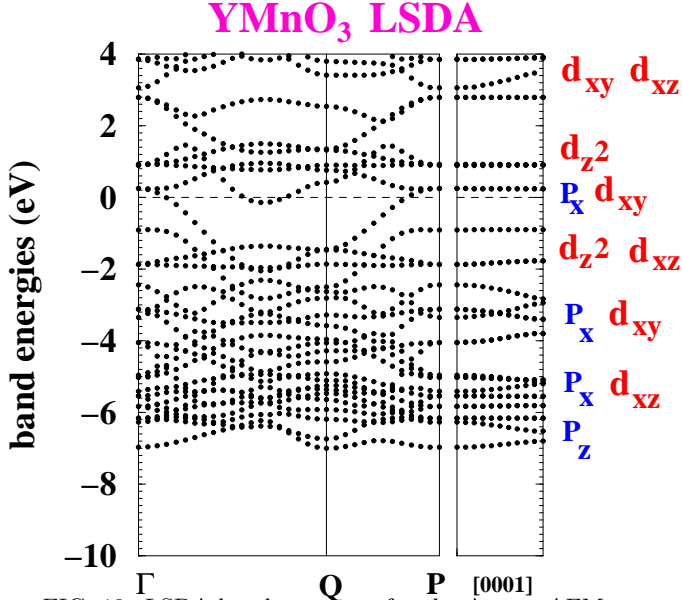}}
\caption{LSDA band structure for the A-type AFM paraelectric YMnO$_3$. The orbital character is given in terms 
of the hexagonal cartesian coordinates. According to this calculation the system is a metal.
\label{ymn}}
\end{figure}

The structure of the paraelectric YMnO$_3$ (P6$_3$/mmc symmetry) is shown in Figure \ref{ymn.fig}. The Mn ions, 
sited on close-packed hexagonal positions, are surrounded by corner-sharing bipyramidal cages of oxygens. 
The unit cell is made of eight 1$\times$1 hexagonal planes enclosing a total of 10 atoms.
In the ferroelectric phase (P6$_3$cm symmetry) the MnO$_5$ bipyramids are tilted around the axis passing 
through the Mn and parallel to one of the triangular base sides, thus the hexagonal planes are
$\sqrt{3}\times\sqrt{3}$ and the unit cell has 30 atoms. However, for the purposes of illustrating the 
effects of the pseudo-SIC, the small oxygen rotations of the ferroelectric structure are not significant, 
thus in the following we only consider the paraelectric phase. 

The technical features of our calculations are the same as those used in Ref.\onlinecite{fh}, where a detailed 
study of density of states, band structure and orbital charges within LSDA can be found. Here we focus 
on the remarkable changes in the band structure of this material produced by the pseudo-SIC.

\begin{figure}
\epsfxsize=9.0cm
\centerline{\epsffile{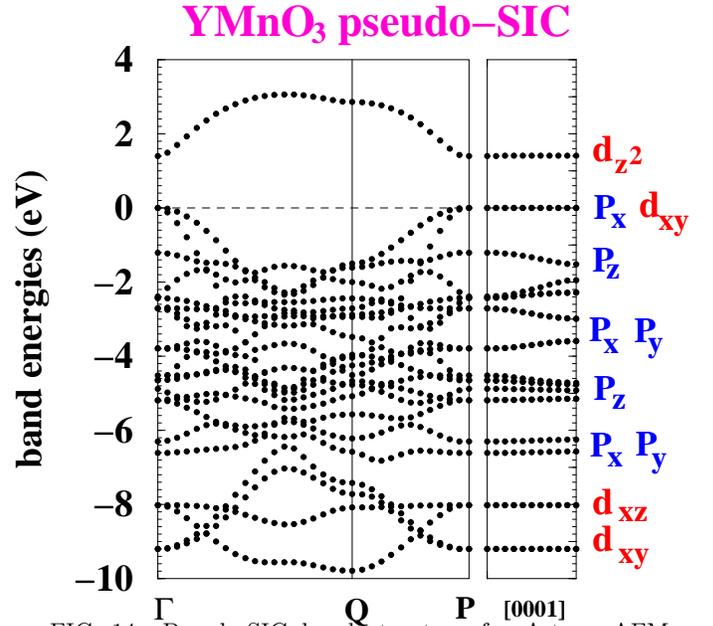}}
\caption{Pseudo-SIC band structure for A-type AFM, paraelectric YMnO$_3$. The pseudo-SIC opens a gap  
between empty d bands and filled pd $\sigma$-hybridized bands.
\label{ymn.sic}}
\end{figure}

An obvious requirement for a ferroelectric material is that it must be insulating, but the LSDA 
calculations describes YMnO$_3$ as a metal, as can be seen in Figure \ref{ymn}. (The hexagonal crystal 
field splitting and the cartesian coordinates are the same as those described previously for MnO and NiO.)
In hexagonal symmetry the four d electrons of the Mn$^{3+}$ ion entirely occupy the two orbital 
doublets ($d_{xy}^{\uparrow}$, $d_{x^2-y^2}^{\uparrow}$) and ($d_{xz}^{\uparrow}$, $d_{yz}^{\uparrow}$),
leaving the $d_{z^2}^{\uparrow}$ orbital, which is the highest in energy, empty. This ordering causes 
a magnetic moment on Mn equal to 3.8 $\mu_B$, slightly lower than its nominal value 4 $\mu_B$
because of Mn d-O p hybridization.  

However the LSDA crystal field splitting is too small to open a gap, and one band crosses $E_F$ in the direction 
$\Gamma$-Q-P which is parallel to the $k_z$=0 plane. This band comes from a mix of Mn $d_{xy}^{\uparrow}$, 
$d_{x^2-y^2}^{\uparrow}$ orbitals and $p_x^{\uparrow}$, $p_y^{\uparrow}$ orbitals from the oxygens lying 
in-plane with the Mn (i.e. it is a pd$\sigma$ hybridization). In contrast, the band energies are flat 
along [0001], as is typical for strongly layered compounds. (Notice that according to our LSDA calculations 
the system is metallic even within the ferroelectric phase.)

The metallicity is a fatal shortcoming since it precludes the possibility of accessing any ferroelectric
properties, such as spontaneous polarization, Born effective charges, and piezoelectric tensors.
(according to our calculations the system is still metallic within the ferroelectric structure)

The pseudo-SIC (Figure \ref{ymn.sic}) repairs the fault: a gap, E$_g$= 1.40 eV, opens (in excellent agreement 
with the experimental value E$_g$= 1.47 eV) between the empty $d_{z^2}^{\uparrow}$ band and the fully occupied 
pd$\sigma$ band. Thus the pseudo-SIC calculations describe the paraelectric YMnO$_3$ as an intermediate 
Mott-Hubbard charge-transfer insulator, in agreement with photoemission experiments\cite{aken} and previous 
LDA+U calculations.\cite{medved} Also, the two filled doublets ($d_{xy}^{\uparrow}$, $d_{x^2-y^2}^{\uparrow}$) 
and ($d_{xz}^{\uparrow}$, $d_{yz}^{\uparrow}$) are shifted down by $\sim$ 3-4 eV with respect to the LSDA values, 
and are located below the O p manifold.

On the basis of the pseudo-SIC calculation which correctly describe the YMnO$_3$ as an insulator we are now
able to study the ferroelectric structural displacements and other properties related to the electric 
polarization. This will be the subject of future work.

\section{Discussion and Summary}
\label{conc}

In this paper we have proposed and implemented an innovative first-principles approach which is able to
improve the LDA and LSDA description of the electronic properties for a vast range of compounds, 
including strongly-correlated and magnetic systems, and requires only a minor increase of computing cost. 
Our strategy, based on the approximation of the true SIC potential with a pseudopotential-like projector, 
was built primarily on the work of Ref.\onlinecite{vkp}, where it was shown that the inclusion of the 
atomic SIC within the pseudopotential construction can consistently improve the agreement of band structures
with photoemission spectra for a range of strongly ionic II-VI and III-V insulators. 
Working on this original idea, we defined a pseudo-SIC functional explicitly dependent 
on the orbital occupation numbers which are self-consistently calculated from the Bloch wavefunctions 
and represent the natural extension to periodic systems of the atomic occupation numbers. 
They can be fractional due to charge hybridization, eigenvalue degeneracy, or Fermi-Dirac distribution.


We tested the pseudo-SIC on three different classes of compounds: non-magnetic semiconductors, magnetic
insulators and ferroelectrics, and found a generally good (sometimes very good) agreement with 
experimental data. In particular, the size and the orbital characterization of the fundamental energy 
gap and the local magnetic moments are much better described than in LSDA.

We expect the same kind of improvement in any system where the electron localization plays a major
role. Examples of systems which could benefit from the pseudo-SIC application are innumerable, including 
bulk systems with d and/or f electrons, Jahn-Teller and orbitally ordered systems, and non-bulk systems 
like defects and impurities, surface and interface states, bulk resonances and core states. 
The SIC may affect a vast range of phenomena, encompassing, among others, surface reconstructions, 
adsorbtion and diffusion of atoms and molecules on surfaces, doping in semiconductors, 
alloys, and homo- and hetero-junctions.

Of course, the discrepancies between theoretical and experimental results which are not attributable
to the SI should not be expected to disappear within pseudo-SIC. A typical example is the fundamental 
energy gap of bulk Si, whose LDA value is $\sim$ 0.5 eV lower than the experimental gap, due to genuine
self-energy effects which are outside the realm of the DFT itself. As a consistency test, we calculated
the band structure of bulk Si within pseudo-SIC. Happily, it comes out to be very similar to the LDA 
band structure. This is a consequence of the fact that each band (either occupied or empty) shares 
the same orbital character (sp$^3$-hybridization), thus the pseudo-SIC mostly produces a trivial shift 
of the whole band manifold. 

Comparing it to other corrective methodologies available in the literature, the pseudo-SIC is closest to the 
spirit of the LDA+U: in both cases the LSDA energy functional is augmented by a term depending on the 
orbital occupation numbers. In the pseudo-SIC the modification is milder, and consists of subtracting 
out a contribution (i.e. the SI) which is already present in the LSDA energy functional. 
The LDA+U is a more radical departure from the LSDA, since it rewrites the whole Coulomb electron-electron 
interaction in terms of local orbitals according to the multiband Hubbard expression (which is, in itself, 
self-interaction free). Based on the undeniably limited set of results presented in this paper, the 
pseudo-SIC seems to be at least as accurate as the LDA+U, but a fair comparison will need a much larger 
body of work.

Finally, we point out that our method suffers a drawback, that is the non-variationality of the
energy functional. This is especially annoying when force or stress calculation is required, since the 
familiar Hellman-Feynman formulations cannot be applied and additional contributions arise due to 
the first-order change of the wavefunctions. We will address these problems and the studies of forces 
and stresses in future publications.

\acknowledgments

We are indebted to Gerhard Theurich for many thoughtful discussions, and to David Vanderbilt
for his critical reading of the manuscript. 
We acknowledge financial support from the National Science Foundation's Division of Materials
Research under grant number DMR 9973076. Also, this work made use of MRL Central Facilities
supported by the National Science Foundation under award No. DMR00-80034. Most of the calculations
have been carried out on the IBM SP2 machine of the MHPCC Supercomputing Center in Maui, HI.

\section{Appendix} 
\label{appen}

\subsection{Review of the USPP formulation}

In the USPP approach\cite{van} the constraint of norm-conservation on the pseudo wavefunctions 
is relieved, so that the pseudo wavefunctions and the associated pseudopotentials can be extremely 
smooth (i.e. ``ultrasoft''). In order to restore the correctly normalized electron charge, 
the charge density is written as:

\startlongequation
\widetext

\begin{eqnarray}
n^{\sigma}({\bf r})\,=\,\sum_{n{\bf k},\sigma}\,f^{\sigma}_{n{\bf k}}\,
\left[|{\psi^{\sigma}_{n{\bf k}}}({\bf r})|^2+
\sum_{\alpha\alpha'}\,\langle \psi^{\sigma}_{n{\bf k}}|\beta_{\alpha}\rangle \,
Q_{\alpha\alpha'}({\bf r})\,\langle\beta_{\alpha'} | \psi^{\sigma}_{n{\bf k}}\rangle\right].
\end{eqnarray}
%

Here $\alpha=[{n,l,m},R]$ and $\alpha'=[{n',l',m'},R]$ are sets of orbital quantum numbers and atomic 
positions R, and $\beta_{\alpha}({\bf r})$ are the USPP projector functions.\cite{van} 
The first term within the square brackets is the ultrasoft charge, and the second term the ``augmented'' charge, 
that is the portion of the valence 
charge which is localized within the atomic core radii and restores the normalization of the 
total charge. The atomic charges $Q_{\alpha\alpha'}$ are, by construction,
\begin{eqnarray}
Q_{\alpha\alpha'}({\bf r})=
\phi_{\alpha}^{AE}({\bf r})\,\phi_{\alpha'}^{AE}({\bf r})
-\phi_{\alpha}^{PS}({\bf r})\,\phi_{\alpha'}^{PS}({\bf r}),
\end{eqnarray}
where $\phi_{\alpha}^{AE}$ and $\phi_{\beta}^{PS}$ are atomic all-electron and
pseudo wavefunctions, respectively. The release of norm-conservation leads to the
generalized Kohn-Sham equations:
%
\begin{eqnarray}
\left(-\nabla^2 + V_{LOC}({\bf r}) + \sum_{\alpha\alpha'} 
|\beta_{\alpha}\rangle\,D_{\alpha\alpha'}^{\sigma}\,\langle\beta_{\alpha'}| +
V_H({\bf r}) + V_{XC}^{\sigma}({\bf r})\right)
\psi^{\sigma}_{n{\bf k}}({\bf r})=
\epsilon^{\sigma}_{n{\bf k}}\,\hat{S}\,\psi^{\sigma}_{n{\bf k}}({\bf r}).
\label{ks}
\end{eqnarray}
\stoplongequation

where $V_{LOC}({\bf r})$ is the local part of the pseudopotential,
$D_{\alpha\alpha'}^{\sigma}$ is the non-local part and $\hat{S}$ is
the overlap matrix which generalizes the orthonormality condition:

\begin{eqnarray}
\hat{S}=\hat{1}+\sum_{\alpha\alpha'}\, |\beta_{\alpha}\rangle\>
q_{\alpha\alpha'} \>\langle\beta_{\alpha'}|.
\end{eqnarray}

Here $q_{\alpha\alpha'}$ are the integrals of the augmented charges
$Q_{\alpha\alpha'}({\bf r})$, and
$<\psi_{nk}|\hat{S}|\psi_{n'k'}>=\delta_{n,n'}\,\delta_{k,k'}$.

Finally, the non-local pseudopotential projector is made up of two contributions:

\begin{eqnarray}
D_{\alpha\alpha'}^{\sigma} = {\tilde D_{\alpha\alpha'}} + \int d{\bf r}\,\left(V_{LOC}({\bf r})+
V^{\sigma}_{HXC}({\bf r})\right)\,Q_{\alpha\alpha'}({\bf r}).
\label{d}
\end{eqnarray}

The first term on the right side of Eq.\ref{d} is the usual
Kleinman-Bylander projector, and contributes to the `bare' pseudopotential
(i.e. it is calculated within the atomic reference configuration).
The second term is specific to the USPP formalism, and represents the 
action which the local and screening potentials exert on the augmented 
charges. Since this term depends on $V^{\sigma}_{HXC}$, it has to be updated
during the self-consistency cycle.

In order to ensure better transferability, two atomic reference states, corresponding to 
different energy values, are usually included in the projector for each angular quantum number. 
As a consequence, the USPP projector contains non-diagonal terms ($\alpha$,$\alpha'$), 
where $\alpha=({\nu}_l,l,m)$, and $\alpha'=({\nu}_l',l,m)$, and $\nu_l$, $\nu_l$'= 1,2.
The atomic reference eigenstates do not need to correspond to bound,
normalized solutions of the free atom, but may be unphysical eigenstates of the Schr\"odinger 
equation, useful to extend the pseudopotential transferability into a larger energy range. 
Thus the atomic pseudo wavefunctions may diverge at large ${\bf r}$, but the projector 
functions $\beta_{\alpha}({\bf r})$ and the matrix ${\tilde D}_{\alpha\alpha'}$ are 
always short-ranged and well-defined by construction.

\subsection{Pseudo-SIC within USPP}

The USPP implementation of the pseudo-SIC requires some generalization of the formalism described 
in Sections \ref{method1} and \ref{method2}. 
The charge densities $n_i^{\sigma}$ of the (pseudo) atomic orbitals $\phi_i$ are:

\begin{eqnarray}
n_i^{\sigma}({\bf r})\,=
p_i^{\sigma}\,\left(|\phi_i({\bf r})|^2+
\sum_{\alpha\alpha'}\,\langle \phi_i|\beta_{\alpha}\rangle \,
Q_{\alpha\alpha'}({\bf r})\,\langle\beta_{\alpha'}|\phi_i\rangle\right)
\label{n_i.uspp}
\end{eqnarray}

and the occupation numbers $p_i^{\sigma}$ become:

\begin{eqnarray}
p_i^{\sigma}=\,\sum_{n{\bf k}}\, f_{n{\bf k}}^{\sigma}\, \langle\psi_{n{\bf k}}^{\sigma}|
\phi_i\rangle \> \langle\phi_i|\psi_{n{\bf k}}^{\sigma}\rangle\times
\nonumber
\end{eqnarray}
\begin{eqnarray}
\left[1+
\sum_{\alpha\alpha'}\,\langle\phi_i|\beta_{\alpha}\rangle\,q_{\alpha\alpha'}\,
\langle\beta_{\alpha'}|\phi_i\rangle \right].
\label{occuspp}
\end{eqnarray}

Furthermore, at variance with the norm-conserving case, the non-local part of the USPP
depends self-consistently on the screening potential itself (see Eq.\ref{d}).
As a result it also must be self-interaction corrected. 
The SI part of the non-local USPP is given by:

\startlongequation

\begin{eqnarray}
\hat{V}_{US}^{\sigma}\>=
\sum_i\,\sum_{\alpha\alpha'}\> |\beta_{\alpha}\rangle\,\left(
{1\over 2}\,p_i^{\sigma}\, 
\int\! d{\bf r}\>V_{HXC}^{\sigma}[n_i^{\sigma}({\bf r});1]
\>Q_{\alpha\alpha'}({\bf r})\right)
\>\langle\beta_{\alpha'}|.
\label{sic2}
\end{eqnarray}

Thus, the pseudo-SIC KS equations finally are:

\begin{eqnarray}
\left[-\nabla^2 + \hat{V}_{LOC} + \hat{V}_{HXC}^{\sigma} +
\sum_{\alpha\alpha'} |\beta_{\alpha}\rangle\,
D_{\alpha\alpha'}^{\sigma}\,\langle\beta_{\alpha'}|
- (\hat{V}_{SIC}^{\sigma}+\hat{V}_{US}^{\sigma})\right]
|\psi^{\sigma}_{n{\bf k}}\rangle=
\epsilon^{\sigma}_{n{\bf k}}\,\hat{S}\,|\psi^{\sigma}_{n{\bf k}}\rangle,
\label{ks-sic_1}
\end{eqnarray}

where $\hat{V}_{SIC}^{\sigma}$ is given by Eqs.\ref{sic1},\ref{rad}, and \ref{rad1}. 

The total energy is the same as that given in Eq.\ref{te_2}, except for the fifth term which now is:

\begin{eqnarray}
\sum_{n{\bf k},\sigma}\>f^{\sigma}_{n{\bf k}}\,
\langle\psi_{n{\bf k}}^{\sigma}\,|\,(\hat{V}^{\sigma}_{SIC}+\hat{V}^{\sigma}_{US})\,
|\,\psi_{n{\bf k}}^{\sigma}\rangle.
\end{eqnarray}

\stoplongequation


\end{multicols}

\begin{thebibliography}{99}
\bibitem{note_cor}
We use the term ``strongly-correlated'' to describe systems whose electron 
charge retains atomic-like features such as strong space localization, poorly-dispersed 
band energies and large on-site Coulomb energies.
\bibitem{note}
The inclusion of the SIC for the core states increases the spatial localization and
the binding energy of the electron core charges. This in turn has two opposite 
consequences on the valence states: it increases the ion-core screening, and reduces 
the repulsion due to the orthogonality conditions. As a net effect, the valence electron 
charge is almost unaffected.
\bibitem{comm1}
In KS theory E$_g$ = $\epsilon_{\mbox{lumo}}$[N+1] -- $\epsilon_{\mbox{homo}}$[N], where $\mbox{lumo}$ 
and $\mbox{homo}$ are the lowest unoccupied and highest occupied electron states, and N is the number of electrons. 
If the $\mbox{lumo}$ is itinerant we have 
$\epsilon_{\mbox{lumo}}^{\mbox{LDA}}[N+1]=\epsilon_{\mbox{lumo}}^{\mbox{LDA}}[N]$, and no SIC is necessary. 
If otherwise the $\mbox{lumo}$ is localized, then 
$\epsilon_{\mbox{lumo}}^{\mbox{LDA}}$[N+1] $>$ $\epsilon_{\mbox{lumo}}^{\mbox{LDA}}$[N], and the SIC should 
be applied to $\epsilon_{\mbox{lumo}}^{\mbox{LDA}}$[N+1] and not to $\epsilon_{\mbox{lumo}}^{\mbox{LDA}}$[N].
\bibitem{reason}
The SIC-LSDA application to the d bands only is motivated\cite{svane} by the fact 
that it gives a lower total energy. However it is shown\cite{af} that this criterion 
is questionable since the sign of the SIC energy depends on the specific form of 
the local exchange-correlation functional used.
\end{thebibliography}
\end{document}